\documentclass[twoside,twocolumn,9pt]{article}
\usepackage[super,sort&compress,comma]{natbib} 
\usepackage[version=3]{mhchem}
\usepackage[left=1.5cm, right=1.5cm, top=1.785cm, bottom=2.0cm]{geometry}
\usepackage{balance}
\usepackage{times,mathptmx}
\usepackage{sectsty}
\usepackage{graphicx} 
\usepackage{lastpage}
\usepackage[format=plain,justification=justified,singlelinecheck=false,font={stretch=1.125,small,sf},labelfont=bf,labelsep=space]{caption}
\usepackage{float}
\usepackage{fancyhdr}
\usepackage{fnpos}
\usepackage[english]{babel}
\usepackage{array}
\usepackage{droidsans}
\usepackage{charter}
\usepackage[T1]{fontenc}
\usepackage[usenames,dvipsnames]{xcolor}
\usepackage{setspace}
\usepackage[compact]{titlesec}
\usepackage{hyperref}
\usepackage[utf8]{inputenc}
\usepackage[english]{babel}
 
\usepackage{epstopdf}

\definecolor{cream}{RGB}{222,217,201}
\usepackage{soul} 
\begin{document}

\thispagestyle{plain}
\fancypagestyle{plain}{\renewcommand{\headrulewidth}{0pt}}

\makeFNbottom
\makeatletter
\renewcommand\LARGE{\@setfontsize\LARGE{15pt}{17}}
\renewcommand\Large{\@setfontsize\Large{12pt}{14}}
\renewcommand\large{\@setfontsize\large{10pt}{12}}
\renewcommand\footnotesize{\@setfontsize\footnotesize{7pt}{10}}
\makeatother

\renewcommand{\thefootnote}{\fnsymbol{footnote}}
\renewcommand\footnoterule{\vspace*{1pt}%
\color{cream}\hrule width 3.5in height 0.4pt \color{black}\vspace*{5pt}} 
\setcounter{secnumdepth}{5}

\makeatletter 
\renewcommand\@biblabel[1]{#1}            
\renewcommand\@makefntext[1]%
{\noindent\makebox[0pt][r]{\@thefnmark\,}#1}
\makeatother 
\renewcommand{\figurename}{\small{Fig.}~}
\sectionfont{\sffamily\Large}
\subsectionfont{\normalsize}
\subsubsectionfont{\bf}
\setstretch{1.125} 
\setlength{\skip\footins}{0.8cm}
\setlength{\footnotesep}{0.25cm}
\setlength{\jot}{10pt}
\titlespacing*{\section}{0pt}{4pt}{4pt}
\titlespacing*{\subsection}{0pt}{15pt}{1pt}

\fancyfoot{}
\fancyfoot[RO]{\footnotesize{\sffamily{1--\pageref{LastPage} ~\textbar  \hspace{2pt}\thepage}}}
\fancyfoot[LE]{\footnotesize{\sffamily{\thepage~\textbar\hspace{3.45cm} 1--\pageref{LastPage}}}}
\fancyhead{}
\renewcommand{\headrulewidth}{0pt} 
\renewcommand{\footrulewidth}{0pt}
\setlength{\arrayrulewidth}{1pt}
\setlength{\columnsep}{6.5mm}
\setlength\bibsep{1pt}

\makeatletter 
\newlength{\figrulesep} 
\setlength{\figrulesep}{0.5\textfloatsep} 

\newcommand{\topfigrule}{\vspace*{-1pt}%
\noindent{\color{cream}\rule[-\figrulesep]{\columnwidth}{1.5pt}} }

\newcommand{\botfigrule}{\vspace*{-2pt}%
\noindent{\color{cream}\rule[\figrulesep]{\columnwidth}{1.5pt}} }

\newcommand{\dblfigrule}{\vspace*{-1pt}%
\noindent{\color{cream}\rule[-\figrulesep]{\textwidth}{1.5pt}} }

\makeatother

\twocolumn[
  \begin{@twocolumnfalse}

\vspace{3cm}
\sffamily
\begin{tabular}{m{18cm} p{0cm} }

\noindent\LARGE{\textbf{Double emulsion drop evaporation and resurfacing of daughter droplet}} &  \\
\vspace{0.3cm} & \vspace{0.3cm} \\

\noindent\large{Muhammad Rizwanur Rahman and Prashant R. Waghmare$^{\ast}$ \footnotetext{waghmare@ualberta.ca}}  & \\

\noindent\normalsize{\textit{interfacial} Science and Surface Engineering Lab (\textit{i}SSELab), Department of Mechanical Engineering, University of Alberta, Edmonton, Alberta T6G2G8, Canada;} & \\
\vspace{0.5cm}

\noindent\normalsize{{\color{black}In this study, we present experimental and theoretical analyses of double emulsion drop evaporation. After the apparent completion of evaporation of the inner phase of a double emulsion drop, surprisingly, a resurfacing of a daughter droplet is observed. We further investigated to hypothesize  this phenomenon  which allowed us to obtain a prolonged fixed contact line evaporation for a single phase drop along with similar occurrence of resurfacing as of the double emulsion drops.}} & 

\end{tabular}
\end{@twocolumnfalse} \vspace{0.6cm}
]

\renewcommand*\rmdefault{bch}\normalfont\upshape
\rmfamily
\section*{}
\vspace{1cm}


\footnotetext{\textit{$^{\ast}$waghmare@ualberta.ca}}



\section*{Introduction} 
{\color{black}The importance of drop evaporation can be identified in numerous applications such as ink-jet printing, coating technologies~\cite{talbot2014inkjet, calvert2001inkjet},self cleaning~\cite{blossey_SelfCleaning}, bio-sensing~\cite{de2011bio} and droplet based micro-fluidics~\cite{chang2006fluidics, li2006pattern}. For the phenomenon being highly sensitive to surface morphology and its chemical composition~\cite{Qli2016} it sparked numerous researchers across disciplines to conduct theoretical and experimental investigations~\cite{man2016ring, Review_2010, erbil2002drop, hu2002_NumExp}.}\newline
{\color{black}Interestingly, the study of evaporation which is crucial for range of applications, from DNA mapping~\cite{wu2004DNA} to chip manufacturing~\cite{dugas2005_Chip}, has not been extended for multi-phase droplets, though that of a single phase droplet has been well analyzed and understood~\cite{man2016ring, Review_2010, erbil2002drop, hu2002_NumExp}. The study of different aspects of multicomponent drop is sparsely  attended. Double emulsion droplets, a simplest representation of multicomponent drop,  is of paramount importance for their potential in several applications staring from  encapsulation technology, drug delivery to the development of micro-nano scale devices~\cite{Weitz_drug,encapsulation_2003,Screening_2004,Weitz_bio_2,LOC_1,LOC_2}.
{\color{black}Recently, evaporation of a drop of transparent mixture of water, ethanol and anise oil - commonly known as `Ouzo drop' was studied where four phases of evaporation were observed~\cite{tan2016evaporation}. This intrigued researchers to look into the previously unexplored evaporation patterns or modes associated with compound or multiphase droplets.}
The necessity of such studies becomes paramount for technologies that facilitate targeted and encapsulated delivery of drug or active reagents ~\cite{iqbal2015double, nakhare1996preparation} where delivery on demand is crucial. Precise control over the disappearance of the outer protective shell can provide a superior passive control on the time stamping.} 

{\color{black}For evaporation of an isolated liquid sphere in an infinite medium, rate of mass transfer follows a linear relationship with radius, as described by Maxwell`s equation, where the diffusive flux is used as an analogy to the electrostatic potential~\cite{maxwell1890}.
Picknett and Bexon~\cite{picknett1977}, in their pioneering work, distinguished between the existence of two modes of sessile drop evaporation, namely, constant contact radius (CCR) or fixed three phase contact line (fixed TPCL) and constant contact angle (CCA) or moving TPCL. The chaotic existence of both modes is often observed, particularly at the end of the droplet evaporation until a visual observation permits to measure the contact angle. The measurable end of the evaporation is always identified by reporting the diminishing contact angle which is difficult and erroneous to report below $5^{\circ}$.} {\color{black}In reality the existence of the liquid thin film with finite volume is always ignored.  
The transition~\cite{orejon2011transition, chen2012transition, chuang2014transition,jansen2015transition,bourges1995} from fixed  to moving contact line occurs when the evaporating flux at the TPCL dominates over the evaporation through the liquid-vapor interface. With attainment of a critical contact angle, droplet perimeter can no longer remain pinned on the substrate and starts slipping.
Thus, the moving TPCL evaporation mode is observed.
This is a consequence of the competition between an intrinsic adhesion force that arrested the contact line motion and an exertion of a force due to evaporation flux that attempts to overcome this barrier by  contracting the droplet through liquid-vapor interface~\cite{orejon2011transition}. The stick-slip or stick-jump at the moving contact line is another observation researchers have reported for drop evaporation in number of situations ~\cite{he2008surface, orejon2011stick, bormashenko2011evaporation}.
The adaptability of the well advanced single phase droplet evaporation theory was studied for drop evaporation on textured non-wetting situations by~\citeauthor{dash2013} \cite{dash2013}. 
Quite recently, intermittent stick-jump mode in drop evaporation has been reported~\cite{dietrich2015stickJump} where the authors further exploited the stick-jump by maneuvering the surface roughness.}

This paper investigates double emulsion drop evaporation; however, the adaptability of existing theory for single phase drop evaporation is carefully scrutinized before comparing it with that of double-emulsion scenario. This required us to experiment with single phase droplets that are involved in double emulsion study with the same drop liquid and substrate combinations.
{\color{black}{While the single phase drop evaporation study is well studied, {\color{black}evaporation of a double-emulsion droplet may pose a complicated scenario since there are new or modified interfaces with an additional liquid-liquid interface and two TPCLs.} For comparing with the double emulsion drop evaporation, certain modifications are necessary that are developed after confirming the validity of the well-established models used in the literature.}}

{\color{black}We observed surprising} appearance of a daughter droplet after the completion of noticeable evaporation of the inner drop. The unexpected occurrence of a daughter drop is coined as `resurfacing' of daughter droplet. Based on the phenomenological evidence of resurfacing of a tiny drop during double-emulsion drop evaporation, an hypothesis is proposed which is validated with single phase drop evaporation where the occurrence of daughter drop is carefully engineered. 

\section*{Experimental methods}
\begin{figure}[!htbp]
\centering
\includegraphics[width=0.45\textwidth]{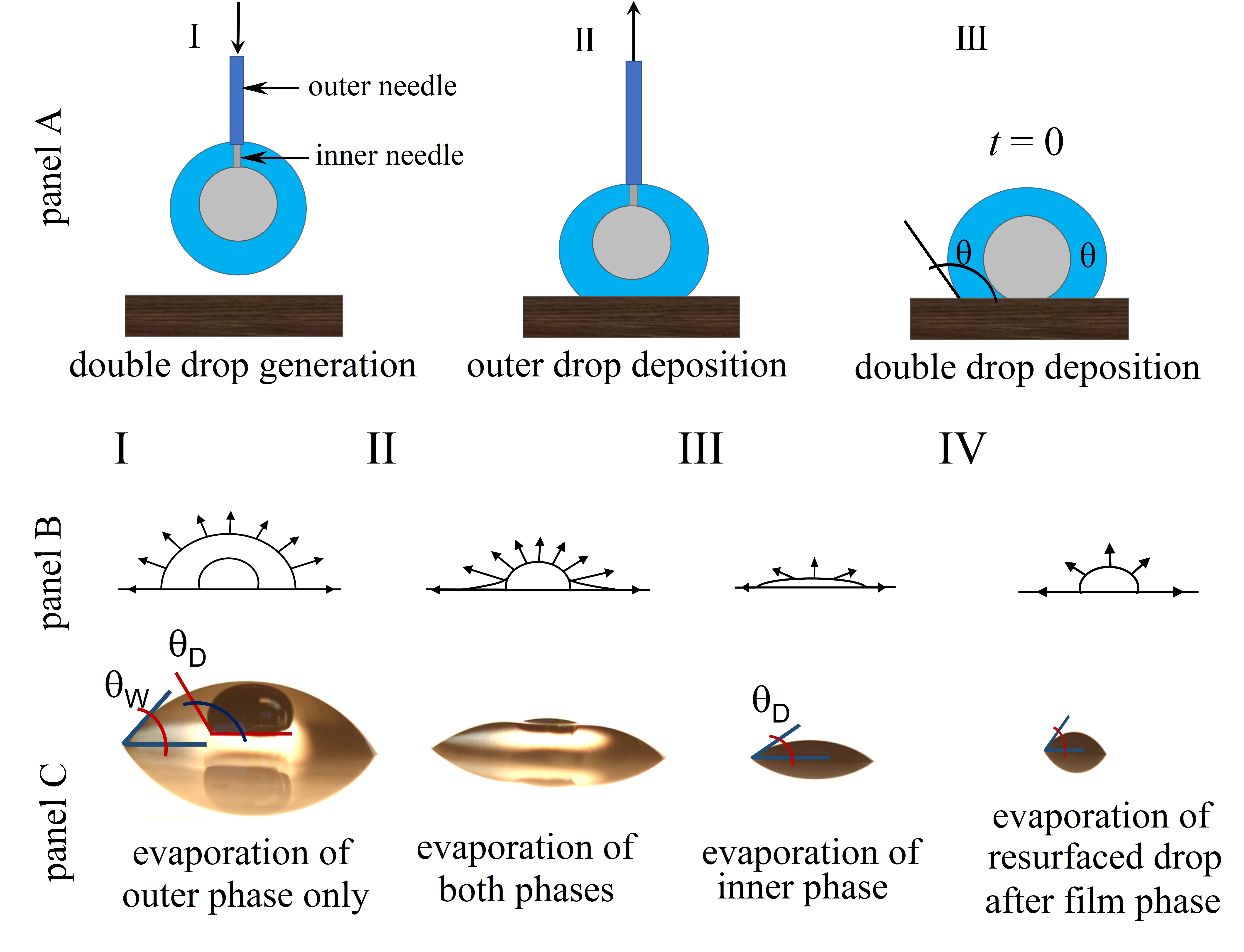}
\caption{{\color{black}Evaporation of  double emulsion drop (images are not drawn to scale): Panel A: (I) generation of double emulsion drop at the tip of the co-axial needle (II) deposition of double emulsion drop (III) evaporating double emulsion drop ; Panel B: Schematic of the double emulsion drop evaporation process:(I) evaporation of the outer phase , (II) disappearing outer phase drop, (III) evaporation of the inner phase (IV) resurfacing of daughter droplet ; Panel C: experimentally observed four stages (I-IV) of double emulsion drops. Synthetic colouring is used to distinguish between inner and outer drop phase.}}
\label{F_Schematic}
\end{figure}

{\color{black} 
\subsection*{Deposition of a double emulsion drop}
Generation of a double emulsion drop required the designing of a customized concentric coaxial needle with two different inlets. With a smaller diameter ($d_o = 0.5 mm$), the inner needle is slightly ($0.5 mm$) protruded outside of the outer needle which has a larger diameter ($d_o = 1.8 mm$).
For the experiments, the inner needle was connected to the deposition unit of DSA 100E (KR$\ddot{U}$SS GmbH, Hamburg, Germany) while a secondary pump was connected to the inlet port of the outer needle.
Initially, the outer phase liquid was pumped through the outer drop inlet port. After the generation of the outer drop of known volume, the inner drop deposition unit of DSA 100E dispensed the inner drop liquid at desired flow rate. The immiscibility between two phases and smaller inner drop volume facilitates the successful generation of the double-emulsion droplet at the tip of the co-axial needle.   
Each experiments were conducted for at least three times. The uncertainty of drop volume was as low as $\pm 0.1 \mu L$.
The generated double-emulsion drop (Panel A of Fig.~\ref{F_Schematic}) at the tip of the co-axial needle was then brought into contact with pre-cleaned substrate to deposit the drop. As the needle was retracted away from the substrates, the drop detached from the needle. Panel B and C schematically identify different distinguished steps of the evaporation.

\subsection*{Droplet liquids and substrates}
An appropriate selection of liquid combination allowed the detachment of the inner drop from needle at the outer drop-air interface. For validation of the proposed modified model for double emulsion drop evaporation, we studied the evaporation of single phase drop of water (saturated vapor concentration, $c_{s} = 0.017~kg/m^3$, diffusion coefficient, $D = 2.4 \times 10^{-5}~ m^2s^{-1}$), diiodomethane ($c_{s} = 0.018~kg/m^3, D = 6 \times 10^{-6}~m^2s^{-1}$) and toluene ($c_{s} = 0.14 ~kg/m^3, D =  8 \times 10^{-6}~m^2s^{-1}$). For double-emulsion drop, diiodomethane was used as inner drop while outer drop was of water. 
Oleophobic substrates ($10~cm\times 4~cm$), pristine adhesive surface ($5~cm\times 5~cm$) and acrylic sheets ($5~cm\times 5~cm$) were used as substrates for evaporation studies. The oleophobic substrates were cleaned with de-ionized water and ethanol prior to each experiment. For cleaning the acrylic sheet, acetylene was used in addition with de-ionized water and ethanol. The average roughness values of the substrates are reported in Table~\ref{T_Substarte}. 

\begin{table}[H]

\centering
{\color{black}
\small
\caption{Surface roughness of the substrates} 
\label{T_Substarte}
\begin{tabular}{llr}

Substrate & Measurement technique & Roughness  \\ 
& (water drop) & (nm)\\
\hline \\
acrylic  & Optical Profilometry & 4-5 \\
oleophobic  & Atomic Force Microscopy & 10-30  \\
adhesive  & Optical Profilometry & 130-135 \\

\end{tabular}
}
\end{table}

\subsection*{Visualization and Contact angle measurements}
After careful deposition on a {\color{black} pre-cleaned substrate}, the drop was allowed to attain equilibrium configuration. The change in contact angles and base diameter were recorded by the in-built imaging system of DSA 100E. The CMOS camera allowed to optimize the image settings and recording of the entire drying period at 60 frames per second from side. An additional synchronized CMOS camera captured the top view of the evaporating drop. 
The outer diameter of the needle was used to calibrate the pixels for their transformation into physical dimension.
Base diameter and contact angles were analyzed from the side view of the droplet, whereas, the top views further assisted in comprehending the process, specially at the end of drying period as discussed later in this article.
To measure the dynamic contact angle from recorded frames, sessile drop technique was employed.
This method utilizes the tangent method for contact angle measurement.Though the Young-Laplace equation fitting method is a better option for measuring static contact angle, but its assumption of symmetric drop shape restricts its applicability for dynamic contact angle measurement. Hence, the tangent method was used for contact angle measurement.

\begin{table*}[!htbp]
\small
  \caption{Contact angles of different drop liquids on the substrates used for experiment}
  \label{T_CA}
  \begin{tabular*}{\textwidth}{@{\extracolsep{\fill}}llllr}
    \hline
    Drop liquid  & Liquid boiling point &  Substrate  & Medium &  Contact Angle \\
      & ($^{\circ}C$) & & &($^{\circ}$), $\pm 3$ \\
    \hline
    water    & 100 & oleophobic   & air   &  $ 80 $ \\
	water	  &	100 & adhesive & air   & $105 $  \\
	water	 & 100 & acrylic   & air   & $78 $  \\
	water    & 100  & acrylic with ring     & air   & $ 85$  \\
	diiodomethane &	181 & oleophobic & air & $ 70  $ \\
	diiodomethane &	181 & oleophobic & water & $ 120 $ \\
	diiodomethane &	181 & oleophobic & water vapor (sat.)& $ 65 $ \\
	toluene & 110.6	& oleophobic & air & $ 38 $ \\
    \hline
  \end{tabular*}
\end{table*}
}

\begin{figure*}[!htbp]
\includegraphics[width=0.98\textwidth]{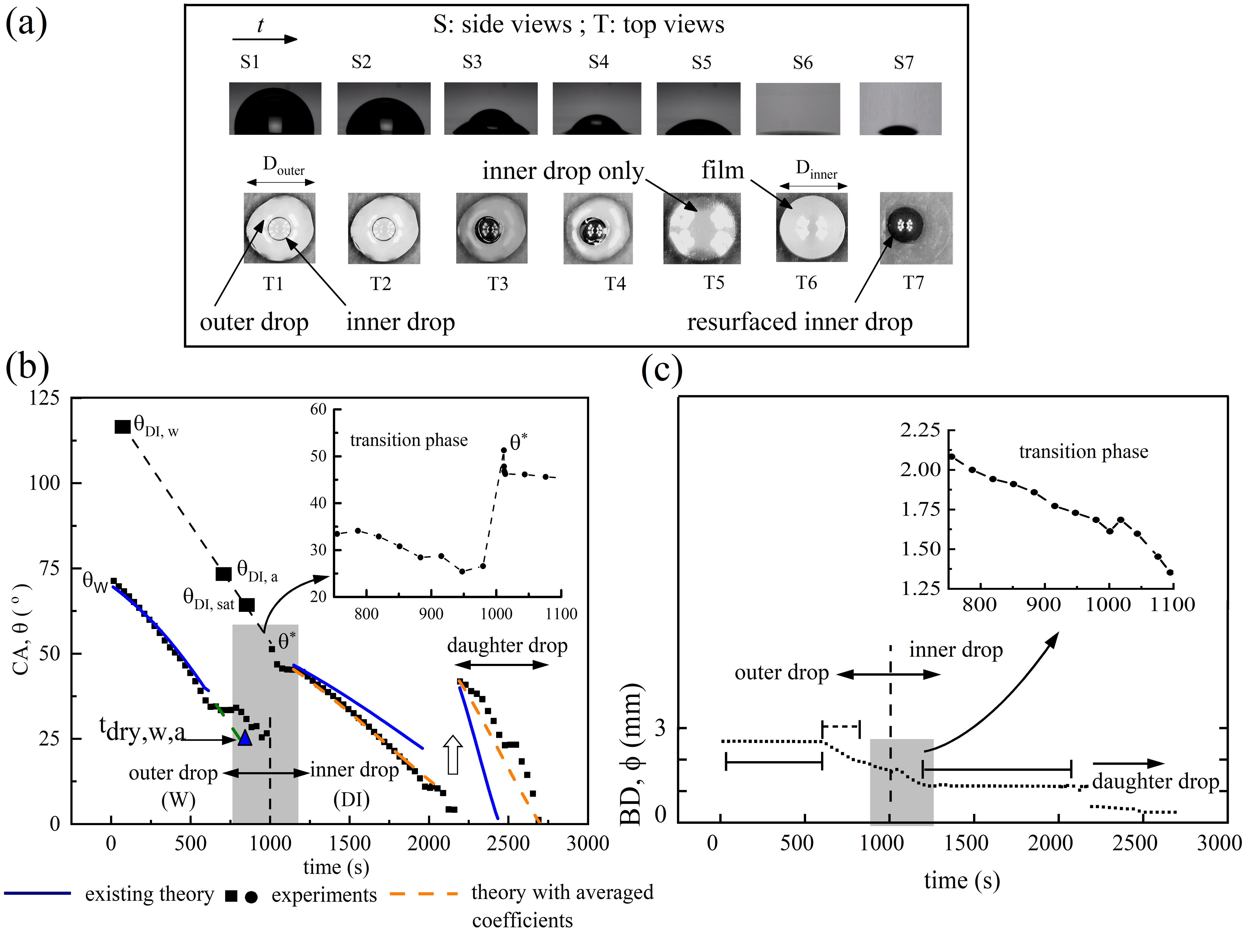}
\caption{Evaporation of water-diiodomethane double emulsion droplet {\color{black}(a) Side views (S1-S4) and corresponding top views (T1-T4) of an evaporating double emulsion drop : evaporation of the outer phase followed by simultaneous evaporation of both the drops in a transition phase (S1-S4 ; T1-T4). Later (S5 ; T5), the outer phase is completely dried and inner drop evaporates until a film phase (S6 ; T6) with nearly zero contact angle is achieved.
S7 and T7 show the resurfacing of the film into a small daughter droplet}.
(b) Dynamic variations in the contact angle  of the three evaporating drops - outer, inner and daughter.   Dynamic variations in the contact angle  of the three evaporating drops - outer, inner and daughter. Symbol representing the experimental observation, dotted and continuous lines are  theoretical predictions with appropriate models.
Inset shows the transition regime where the inner drop attains $\theta^{*}$ - it is different from ($\theta_{DI,a}$), ($\theta_{DI, w}$) or ($\theta_{DI,sat}$) on the same substrate. (c) Corresponding change in the base diameter depicting the same phenomenon magnified in the inset.}
\label{F_Double}
\end{figure*}

\section*{Results and discussion}

\subsection*{Double emulsion drop evaporation}
{\color{black} The so-called double or multiple emulsion drop system, also often termed as emulsion of emulsion, duplex emulsion, multiple emulsion or compound drops interchangeably~\cite{muschiolik2017double}, can be defined as a drop completely engulfed or encapsulated by another immiscible liquid drop~\cite{stone1990breakup}.
}In this study, a diiodomethane drop is encapsulated inside a water drop and evaporation is studied on a oleophobic substrate. Similar to a single phase drop, evaporation starts with fixed TPCL mode for double emulsion drop when the base radius remains unaltered for a considerable time. This consequences in the decrease of contact angle as well as drop height as shown in Fig.~\ref{F_Double}(a) (Side views - S1, S2 ; Top views - T1, T2).
Ensuingly, the outer phase water drop height decreases to that of the inner drop. At one point of time, the two interfaces interact with each other.
We have termed the time period $-$ from this interface interaction to the complete exposure of the inner droplet to air $-$ as `transition regime'. Since the air-water interface is shrinking due to evaporation, eventually the inner drop gets exposed to the air by forming an air-diiodomethane interface.
The partially exposed inner drop is shown in S3 and T3 of Fig.~\ref{F_Double}(a).

It is worthwhile to notice that the second mode of evaporation (moving TPCL) for outer drop is obstructed and shortened by the existence of the inner drop.
The time required to complete the fixed TPCL evaporation mode for a single phase water drop is denoted as $t_{dry, w,a}$ in the figure.
This is longer in comparison with the time observed for same sized water drop cushion in case of double-emulsion drop case.
With this observation, one can argue that the majority of the liquid can be evaporated with a fixed TPCL by carefully maneuvering the outer to inner drop volume or contact line radius ratios.
In the considered volume ratio of water and diiodomethane drops, for outer drop a very short period of a moving TPCL mode was observed. Once both the liquids compete for the evaporation, a third mode was observed with moving TPCL with changing contact angles, where the outer drop merely existed at the bottom of the inner drop as seen in S4 and T4 (transition regime). The water drop forms a precursor film around the inner diiodomethane drop.
A change in the contact angle confirms the visible drying of the water, i.e., outer phase which is shown in the inset of Fig.~\ref{F_Double} (b). The  change in the base diameter suggests the coexistence of water and diiodomethane in liquid-vapor phase along the TPCL. This coexistence of both phases and corresponding changes in the diameter and contact angle is certainly a function of diffusivity, evaporation flux and all other evaporation dynamics parameters. The detailed parametric analysis is required to comment on the importance of these parameter which needs further attention.

After the complete outer drop evaporation, the inner drop liquid-vapour is exposed to the air and thus the dimensions of the inner drop starts to change. The dotted line at $\sim 1000$\textit{s} demarcates the inner drop and outer drop evaporation regime. Once the inner drop was appeared to be completely evaporated the contact angle was almost immeasurable. Most of the studies reported the end of the evaporation up to this phase. But in the case of double-emulsion droplet, resurfacing of a daughter droplet from the thin film is noticed which further evaporates at a different rate.   

Quiet interestingly, for complete inner drop evaporation, the fixed TPCL mode is observed as shown in the Fig.~\ref{F_Double} until the sudden appearance of a daughter droplet.
Hence the apparent completion of evaporation was deceiving due to the limitations of the imaging system which generally allows to measure contact angles as low as $2 - 5^{\circ}$.
We further quantified the evaporation of this resurfaced new drop which is marked as daughter droplet evaporation regime in fig.~\ref{F_Double} (b) and (c).
This emergence of daughter drop from an invisible (as viewed from the side) thin film motivated us to analyze the evaporation from the top view as depicted in fig.~\ref{F_Double} (a).
Despite the fact that the contact angles measured from side view (S6) indicate complete evaporation, the top view shows the presence of a thin-film (T6).
After a few seconds, a daughter drop appears for which the contact angle and base diameter can be measured until it dries out. We have confirmed this observation three times and one can assure that in case of the water-diiodomethane double-emulsion drop evaporation, the diiodomethane drop gets pinned during evaporation.
The pinning of the contact line might be a result of the presence of water vapor along the TPCL which does not allow the drop to change the mode of evaporation from fixed TPCL to moving TPCL. 

{\color{black} In case of fixed TPCL, with decreasing contact angle, the pinning force gradually increases and the drop experiences stronger resistance against contact line movement~\cite{wang2015molecular}. This consequences in the formation of such thin film and at one point the competition between pinning force and the inward contracting force, reported as snapping \cite{de2004capillarity} reach their maxima and suddenly slippage occurs. Such a  slip is not a rare occurrence for drop  evaporation~\cite{dietrich2015stickJump}. When a droplet evaporation shows alternating switch between fixed and moving TPCL mode, the reconfigured drop demonstrate the jump in the contact angle. But in the situation considered here, pinning of TPCL  is not a random occurrence, it is prolonged until the end of the fixed contact line evaporation which is very end of drying period. This delayed fixed TPCL mode with significantly higher evaporation at the contact line of pinned thin film results in a quicker slip motion followed by a formation of daughter drop with a moderately larger contact angle. Non-axisymmetrical snapping off of the contact line can be attributed to the nonuniform surface roughness at the contact line.}
The observed evaporation on outer, inner and daughter droplet is compared with existing models and modifications were required to obtain the appropriate theoretical predictions. Hence, prior to propose modification to the well-established approaches, confirmation of the validity of considered theoretical models for single phase drops is of paramount importance. In the latter section we have validated the theoretical model with that of the single phase drop evaporation using similar drop-substrate configuration. This is then further extended to double emulsion case. 
\subsection*{Single phase drop evaporation}
\begin{figure}[!htbp]
\centering
\includegraphics[width=0.48\textwidth]{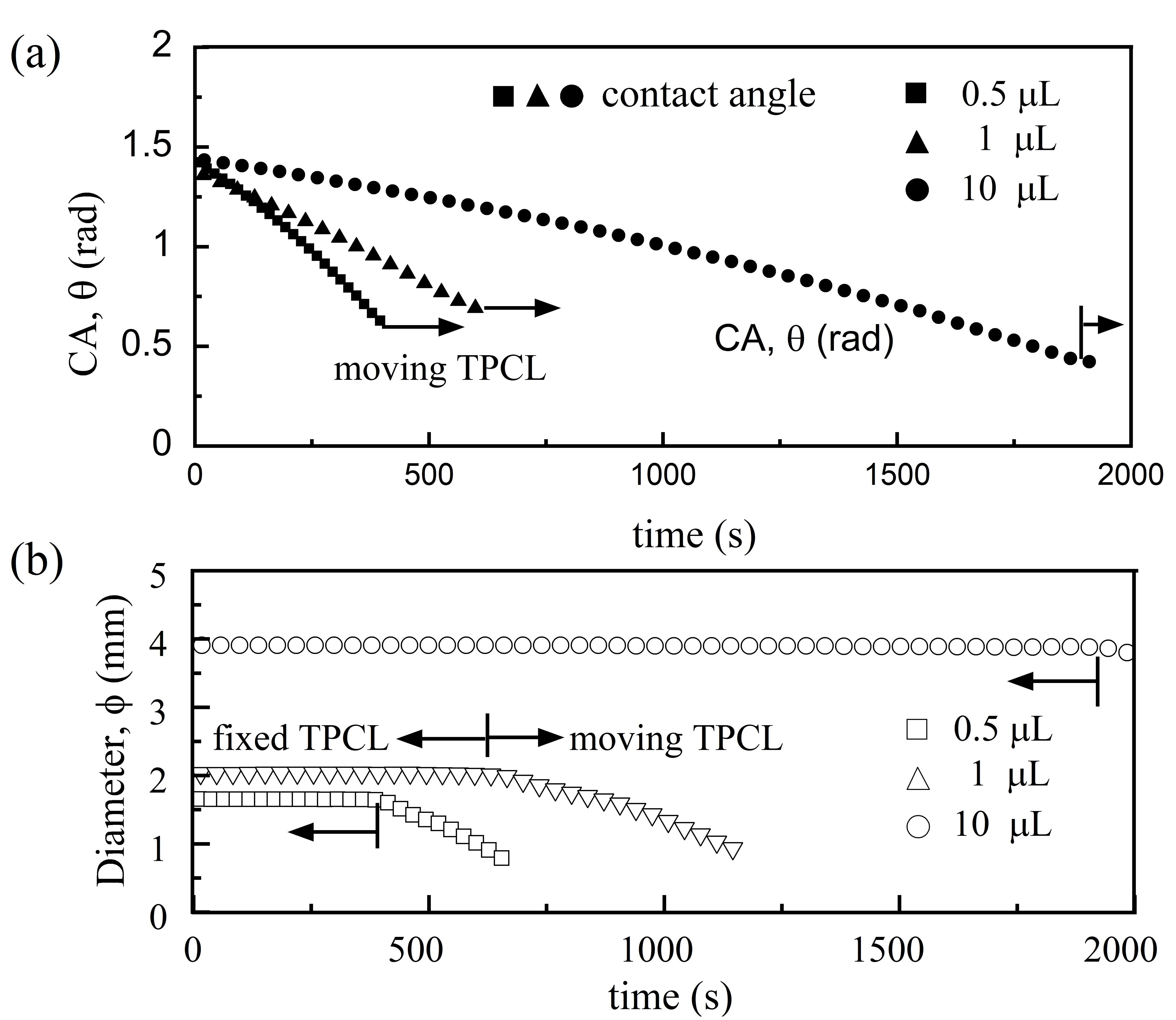}
\caption{Evaporation of single phase droplets:
(a)Contact angles and (b) base diameter of evaporating water droplets of different volumes demonstrate both fixed and moving TPCL during the drying time.}
\label{F_Single}
\end{figure}

\begin{figure*}[!htbp]
\centering
\includegraphics[width=0.98\textwidth]{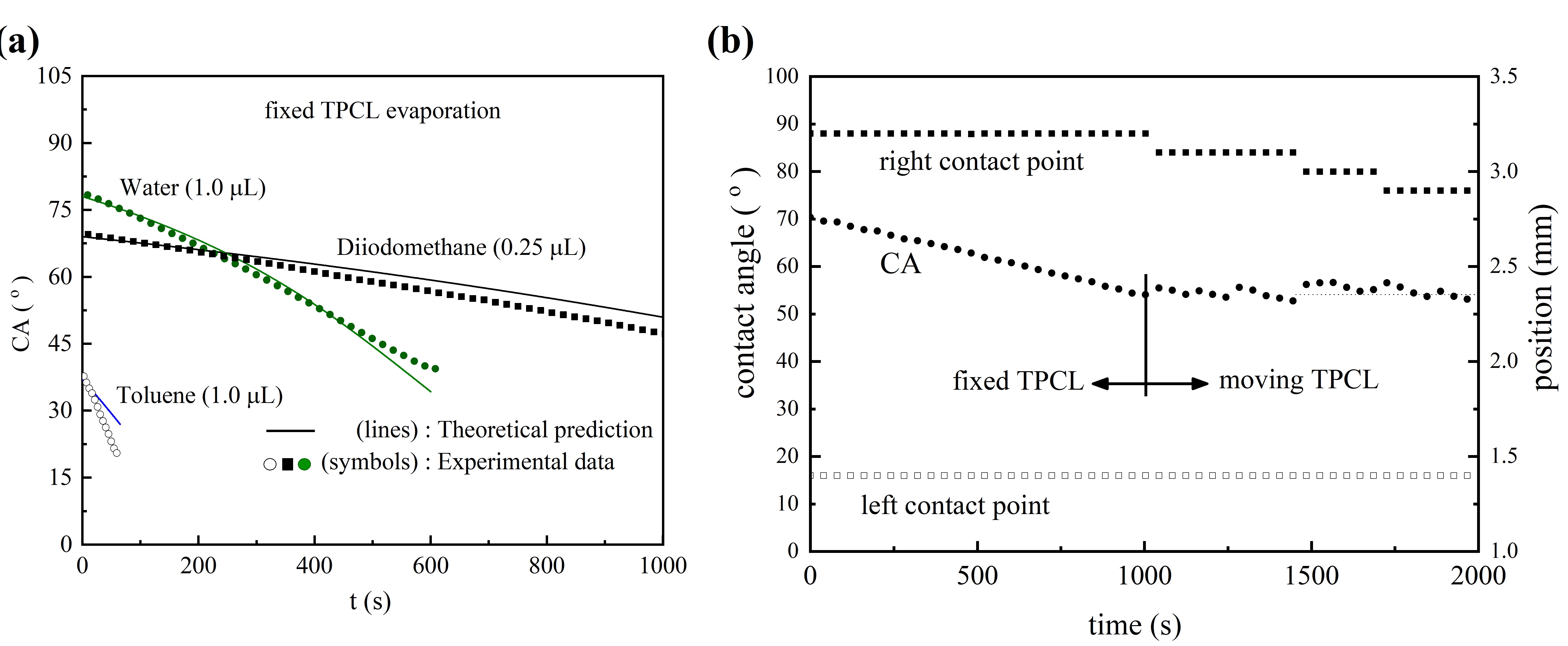}
\caption{(a) Comparison between experimental data and theoretical model~\cite{dash2013} for fixed TPCL mode for single phase droplet evaporation (b) evaporation of a diiodomethane droplet $-$ fixed TPCL mode for the first half of the drying time is followed by intermittent moving TPCL mode with a number of stick-slips.}
\label{F_Single_2}
\end{figure*}

\begin{figure}[!h]
\centering
\includegraphics[width=0.48\textwidth]{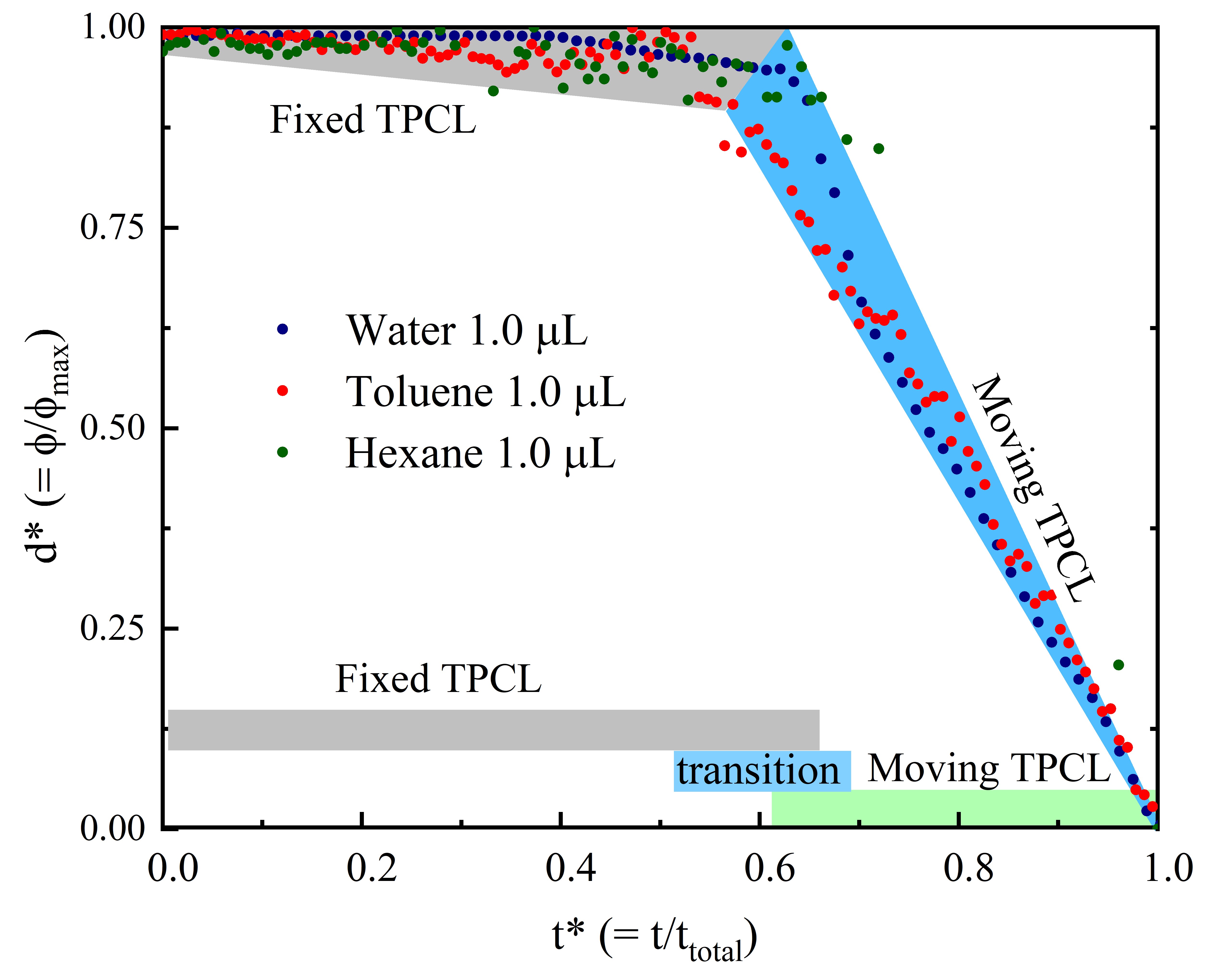}
\caption{Droplets of liquids with varying volatility follows fixed TPCL mode more than half of its drying period and the onset of transition between two modes of evaporation occurs approximately at same time fraction.}
\label{F_Single_3}
\end{figure}

{\color{black}At first, the water (outer)) and diiodomethane (inner) drops of different volume were analyzed separately which will be further compared to the double emulsion drop scenario.} As in Fig.~\ref{F_Single} (a) and (b), water droplets of different initial volumes are seen to evaporate~\cite{mchale2005mixed, dash2011mixed} in a similar fashion. Theoretical estimations for fixed TPCL mode, discussed later in this paper, are compared with experimental observations as presented in Fig.~\ref{F_Single_2} (a) for a range of droplet volumes with varied volatility and diffusion property.

A special attention is given to the complete evaporation of a $0.25 \mu L$ diiodomethane droplet since the inner diiodomethane drop for double emulsion drop evaporation study is of same volume. Figure~\ref{F_Single_2} (b) suggests that in case of single phase diiodomethane drop the number of `stick-jump'~\cite{dietrich2015stickJump} scenario is observed.
To pinpoint this observation, the position of two ends of the base diameter is traced as depicted in Fig.~\ref{F_Single_2} (b).
{\color{black}Similar observations were made for larger ($1 \mu L$) diiodomethane drop except the fact that the increase in volume increases the drying time. Larger contact angle hysteresis is one of the factors that causes such `stick-jump' behavior~\cite{dietrich2015stickJump}. Hence, hysteresis was measured by tensiometer (K100, KR$\ddot{U}$SS GmbH, Hamburg, Germany) and  goniometer (DSA100E, KR$\ddot{U}$SS GmbH). The contact angle hysteresis for oleo-phobic substrate, acrylic sheet and adhesive surface were $36^\circ$, $30^\circ$ and $54^\circ$, respectively. The high contact angle hysteresis contributes to the dominance of fixed TPCL mode~\cite{kulinich2009effect}}.
Prior to discussing the theoretical aspect of the evaporation, it is worth mentioning that
the time fraction required to complete the first mode (fixed TPCL) of evaporation of liquids with different volume and volatility falls around the same time range. In Fig.~\ref{F_Single_3}, for all liquid-solid combinations considered in this study, a fixed TPCL mode of evaporation is observed for a time period more than half of the total drying time.

{\color{black}
For comparing these results with double emulsion drop case,  a single phase droplet model is adapted. For evaporation of a sessile droplet, contact line dynamics and surface morphology complicate the scenario where contact angle as well as base diameter of the TPCL~\cite{birdi1989, picknett1977} dictate the dynamics.}
The diffusion model proposed by Popov~ \cite{popov2005} for a sessile droplet (of mass $M$ and density $\rho$) with contact radius, $R_c$ and contact angle, $\theta$, takes the form of equation \ref{Eqn_2} :
\begin{equation}
	\frac{dM}{dt} = \rho \frac{dV}{dt} = - \pi R D \Delta c f(\theta) 	
	\label{Eqn_2} 
\end{equation} 
Here, the function $f(\theta)$ is given by the following expression 
\begin{equation}
	f(\theta) = \frac{sin \theta}{1+ cos \theta} + \int_o ^\infty \frac{1+cosh (2\theta \tau)}{sinh (2 \pi \tau)} tanh [(\pi - \theta) \tau] d\tau
\label{Eqn_3} 
\end{equation}
where, $\tau$ is non-dimensional drying time~\cite{popov2005}. \newline
As the droplet gets pinned during fixed TPCL mode, the loss in mass translates into corresponding decrements in height and contact angle until a critical contact angle is attained.
Though a droplet would like to evaporate without any additional penalty in its energy, by maintaining equilibrium contact angle, the pinning of the TPCL and dominant evaporation flux across the liquid-air interface create resistance against the smooth decrease in base diameter \cite{shanahan1995EnergyBarrier}.
But as a critical angle is approached, the evaporation flux at the TPCL becomes large enough to surpass the energy barrier resulting in the change of base diameter. Occasionally, in this second mode of evaporation the stick-slip~\cite{anan2009} or stick-jump behaviour of TPCL is noticed\cite{dietrich2015stickJump}. 
Assuming the spherical cap assumption along with the functional variation of contact angle (Eqn \ref{Eqn_3}), the instantaneous change of droplet contact angle in moving TPCL with fixed contact angle mode can be derived as~\cite{dash2013}: 
\begin{equation}
\label{Eqn_5}
\frac{d \theta}{dt} = -\frac{D \Delta c}{\rho R^2} (1+ cos \theta)^2 f(\theta) 
\end{equation}

\noindent The estimations, predicted by Eq.~\ref{Eqn_5}, are compared with experimental data in Fig.~\ref{F_Single_2} (a), i.e., for the single phase droplet scenario, which clearly suggests that the selected theoretical model~(Eqn.~\ref{Eqn_5}) can predict the droplet evaporation dynamics. However, one can detect deviation from theoretical predictions for highly volatile liquid drop as noticed for toluene in Fig.~\ref{F_Single_2} (a).  

\subsection*{Double drop's disparity from single drop}

Following to the validation of single phase, the similar model was extended for all three drops (outer, inner and daughter) of double emulsion droplet. For a double emulsion droplet, one might expect that after the complete outer drop evaporation, the inner drop attains the Young's configuration. The inner diiodomethane drop attains the equilibrium inside water medium with contact angle $\theta _{DI,w}$ and by the time it is exposed to  air, contact angle reduces to $\theta^*$. For analyzing the significance of $\theta^*$ the contact angle of the diiodomethane in saturated water vapor ($\theta_{DI,sat}$)as well as in water medium ($\theta_{DI,w}$) is also measured and compared. The contact angles of diiodomethane in air, water medium and in saturated water vapor ($\theta_{DI,sat}$) are provided in Table~\ref{T_CA} and Fig.~\ref{F_Double}(b). The new configuration with contact angle, $\theta^*$ less than $\theta_{DI,air}$ and $\theta_{DI,sat}$, suggests the change in the local surface energy of the solid, i.e., solid-air interfacial energy. The transition from $\theta _{DI,w}$ to $\theta^*$ is due to the evaporation of the outer drop and the appearance of new configuration compared to $\theta_{DI,air}$ and $\theta_{DI,sat}$ might be due to the adsorption of water as well as diiodomethane vapor. We assume, this change follows similar behaviour as $\theta_{w}$, hence in Fig.\ref{F_Double}(b) we connect $\theta _{DI,w}$ to $\theta^*$ with a dashed line parallel to the evaporation of single phase water drop evaporation passing through $\theta _{DI,a}$ and $\theta _{DI,sat}$. 
As observed for saturated environment wettability studies~\cite{Jacobi_impact}, the saturated vapour not only maintains the contact angle closer to theoretically predicated Young's angle, but also circumvent the decrements in the contact angle due to evaporation~\cite{Jacobi_impact,ismail2018optical}. 
During the outer water drop evaporation the surrounding medium for the inner drop gets saturated with the water vapour which further gets adsorbed on the solid surface.
Once the outer liquid cushion is evaporated, inner drop is suddenly exposed to a modified surface energy interface that results in another marginal decrease in the contact angle, $\Delta \theta \sim 5^\circ$ (inset of Fig.~\ref{F_Double}(b)) with sudden increase in  base diameter, $\Delta \Phi \sim 0.1~mm$ (inset of Fig.~\ref{F_Double}(c)) of the inner drop.\\

It is important to comment on the theoretical modeling of evaporation, validated in Fig.~\ref{F_Single} (b), for double-emulsion droplet case, in particular, for inner and resurfaced daughter drop.
It is evident that, double-emulsion drop evaporation follows the same modeling as of single phase until the onset of the transition regime.
One can argue that the role of the inner drop is negligible for outer drop evaporation, hence the continuous line, the behavior predicted for a single phase drop, perfectly matches with the experiential results.
If we consider a single phase water drop (without inner drop) of the total volume equal to the volume of double emulsion drop, the fixed TPCL evaporation in air can be observed up to $t_{dry,w,a}$ as shown in the Fig.\ref{F_Double} (b). But the presence of the inner drop alters the total evaporation time as well as the mode of the evaporation.
Since the evaporation of inner and daughter drop is mainly of fixed TPCL, it is worthwhile to validate the single phase drop evaporation models for inner and daughter drop.
As presented for outer drop evaporation, the continuous lines in Fig.~\ref{F_Double}(b) represent the theoretical behaviour predicted by Eq.~\ref{Eqn_5} for inner and daughter drop. It is evident that this approach either over predicts or under predicts for inner and daughter drops. We carefully performed the parametric analysis and concluded that the presented modelling is sensitive to the concentration and diffusion of the phases involved in the evaporation.
For a double-emulsion drop evaporation case, it is debatable whether outer or inner drop properties play a role or combined properties need to be considered. The tuning of the theoretical model suggests that while considering the model for inner drop and daughter drop evaporation, volume weighted averaging of the concentration gradient and the diffusion coefficients predicts the behaviour closer to the experimental observations. The dashed lines in Fig.~\ref{F_Double} (b) depict the modified theoretical predictions with appropriate averaged properties of liquids.

\begin{figure*}[!htbp]
\includegraphics[width=0.98\textwidth]{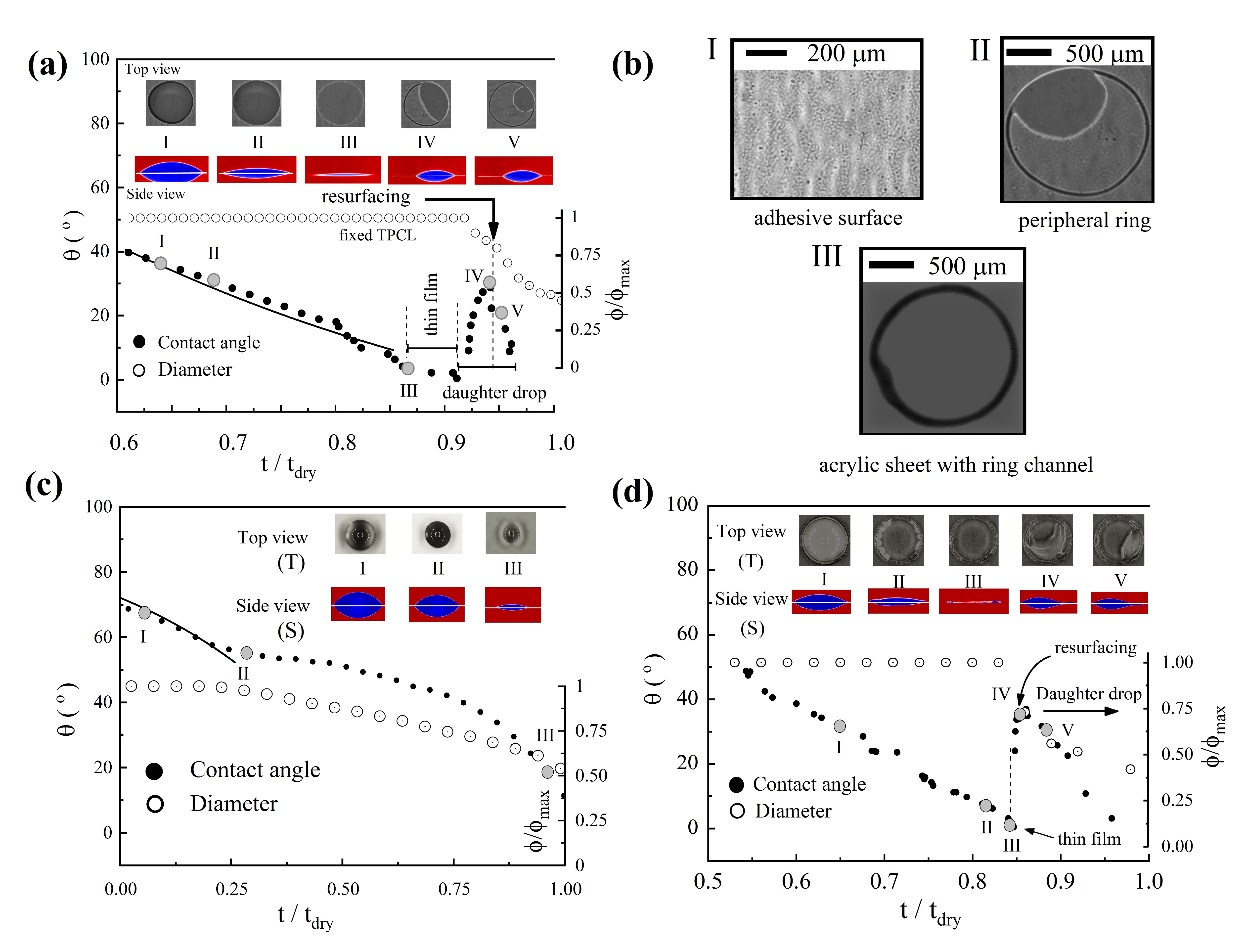}
\caption{{\color{black}Evaporation and resurfacing of droplet. (a) Evaporation on adhesive surface incorporates thin film phase and resurfacing into a smaller droplet. The grey circles with numbers refer to the corresponding top and side views. Experimental data for contact angles and base diameter (non dimensional) are presented with filled and empty symbols. (b) I $-$ microscopic view of the adhesive surface shows the micro-nano features of the surface (b) II $-$ a ring like impression creates local pinning sites along the TPCL (b) III $-$ a micro metric ring is machined on an acrylic sheet to dummy the ring effect (c) water drop evaporation on a acrylic sheet without ring shows no film phase or resurfacing (d) with the ring on the acrylic sheet distinct thin film phase followed by resurfacing of a daughter droplet is seen.}}
\label{F_Resurface}
\end{figure*} 
\noindent This modification in the diffusion and concentration is attributed to the presence of outer phase, i.e., vapor phase of the outer liquid in the vicinity of the TPCL and the liquid-air interface. This might have altered the properties dominating the phenomenon; hence using only single phase properties (continuous lines - blue) over or under predict the experimental observations. This signifies the role of altered surrounding conditions due to the evaporation of two different liquids. Proper quantification of this alteration and the physical explanation may interest researchers for a detail study. {\color{black}Further, the relative solubility of the two associated phases is another aspect which needs to be considered. But in the present study, we have not varied the relative solubility which would be another interesting aspect to investigate.}\newline
{\color{black} It is worthwhile to note that the evaporation experiments were conducted for different volumes of the inner and outer drops ($0.25$ and $0.50~\mu L$ inner diiodomethane drop in $2 , 3$ and $5 \mu L$ outer water drop). Similar evaporation patterns can be observed with the exception of the evaporation rate which depends on the initial volume and contact angle as depicted in Fig.~\ref{F_Single} (a) for single phase drop evaporation.} After proposing the revised model for double emulsion droplet evaporation, the occurrence of daughter droplet is investigated in the upcoming section. 

\subsection*{Resurfacing of evaporating drops}
Interestingly, the  occurrence of daughter drop  was absent in the case of single phase scenario, i.e., diiodomethane drop on the same substrate. It is important to investigate why such a resurfacing was never observed in the case of single phase droplet evaporation.
Therefore, experiments with evaporating water droplet on a number of substrates including acrylic, copper, aluminum sheets, micro textured and adhesive surfaces were performed.
Surprisingly, only adhesive coated surfaces demonstrated the resurfacing of water drops as can be seen in Fig.~\ref{F_Resurface} (a). 
The top two panels depict the top and side views of the evaporating drops.
Change in contact angle with corresponding base diameter is shown with filled and empty symbols, respectively. 
The dominance of fixed TPCL evaporation mode convinced us to conclude that if fixed TPCL evaporation can be significantly prolonged over the drying time, one can observe the resurfacing of the daughter droplet.
Fig.~\ref{F_Resurface} (a) mainly focuses on the end of the fixed TPCL evaporation until it reaches the smallest measurable contact angle.
Different stages presented in the panel are denoted along the change in the contact angle with roman numbers (I - IV). Careful microscopic observation of the adhesives layer suggests that the surface contains micro-nano features (Fig.~\ref{F_Resurface} (b) - I). Thus the surface facilitates the pinning of the contact line which eventually forces the drop to form a film before the daughter drop formation.
In this case, while the daughter drop resurfaces, a big jump in contact angle (from $0$ to $~30 ^{\circ}$) is noticed as shown in Fig.~\ref{F_Resurface} (a).

To view the film and resurfacing of the drop, the camera viewing angle was slightly tilted $(\sim 2^\circ)$ which demonstrates thin film (III and IV in side view) corroborating the presence of liquid film in corresponding top views. The pinning of the TPCL can be confirmed by ring like impression similar to `coffee stain ring' as shown in Fig.~\ref{F_Resurface} (b) - II.
This ring acts as a peripheral pinning location that holds the droplet until it converges to a thin film with vanishing contact angle.
However, it is well established that, if the evaporating flux at TPCL is significantly larger than the evaporation flux across the liquid-air interface, it surpasses the pinning strength and hence, moving TPCL evaporation can be observed. 
In case of the evaporation on adhesive surface, the evaporation flux at TPCL is not large enough until the drop attains the form of a thin film. The moment the evaporation across the air-liquid interface of thin film is not dominant enough, resurfacing triggers into the formation of daughter drop as shown in Fig.~\ref{F_Resurface} (a) IV $-$ VI.
However, a double emulsion or a single diiodomethane drop do not exhibit such behavior on this particular substrate emphasizing on the dependency or sensitivity of this phenomenon on surface-liquid combinations.

With the observation of resurfacing, we identified a critical aspect that dictates the formation of the daughter droplet, i.e., pinning of the contact line for entire evaporation of the droplet.
To validate this proposed hypothesis, we artificially engineered a physical barrier by engraving a ring on an acrylic substrate.
This artificial ring of $1.5mm$ diameter is of the same dimension as that of the base diameter of water drop of a given volume. The top view of the engraved acrylic substrate is shown in Fig.~\ref{F_Resurface}~(b) -III.
For comparison, we initially studied the water drop evaporation on an acrylic substrate without any ring as shown in Fig.~\ref{F_Resurface}(c) which clearly demonstrates the usual modes of the evaporation.
Top and side views at three different time instants also depicts the movement of the TPCL.
Since there is no pinning of the three phase contact line, we cannot expect a thin film phase and subsequent resurfacing.
Discordantly, when evaporation of water drop is observed on the same substrate with a ring (with micro metric depth of $\sim 100\mu m$), we observe the prolongation of fixed TPCL mode over almost the entire drying period as shown in Fig.~\ref{F_Resurface}~(d).
The drop remains pinned along the TPCL (S I, S II ; T I, T II) until the contact angle reaches zero and forms a thin film (T III; S III).
This is followed by the resurfacing of a daughter droplet (S IV, S V ; T IV, T V) as hypothesized, with a jump in contact angle as depicted in the plot along with associated decrease in base diameter.
Thus, by employing our hypothesis \textit{i.e.,} forcefully pinning the TPCL, resurfacing of a daughter droplet is demonstrated on a regular substrate which otherwise doesn't behave similarly.

\section*{Conclusions}
The evaporation of single phase drops well agree with existing theoretical model, however significant deviations have been observed for the double drop case. A modified theoretical approach agrees with the observed evaporation modes for the double emulsion drops. Evaporation of such droplet exhibits the commonly observed modes of evaporation with two new regimes in its drying time. The transition regime from outer to inner drop constitutes a sudden spreading of the inner droplet which results in a wetting scenario that is different from the theoretically expected equilibrium configuration for similar liquid-solid-vapour combination.
The sudden change in the contact angle imprints the complete drying of the outer drop liquid and can be attributed to complete exposure of the inner droplet to environment.
A resurfacing of a daughter droplet is witnessed after the commonly identified completion of the evaporation.
This observation is critically investigated and attributed to the pinning of the three phase contact line. Later, we forcefully pinned the three phase contact line of a single phase droplet by carefully engineering a substrate and a mechanism of daughter droplet resurfacing from thin film is established.  

\section*{Conflict of interest}
The authors declare no conflicts of interest.

\section*{Acknowledgment}
The authors thank Natural Sciences and Engineering Research Council (NSERC) for the financial support in the form of Grant No. RGPIN-2015- 06542. The authors highly acknowledge Dr. Xuehua Zhang for her suggested edits to enhance the representation. We also thank Dr. Aleksey Baldygin for his valuable support in fine tuning the experiments.


\bibliography{Evaporation_Library2.bib} 
\bibliographystyle{rsc} 

\end{document}